\documentclass[preprint]{revtex4-1}

\usepackage{amsmath,amssymb}
\usepackage{graphicx}%
\usepackage{dcolumn}%
\usepackage{bm}%

\begin{document}

\title{Temperature dependence of electronic heat capacity in Holstein model}

\author{N.S.~Fialko}
\email[]{fialka@impb.psn.ru}
\affiliation{Institute of Mathematical Problems of Biology RAS, Pushchino, Russia}

\author{E.V.~Sobolev}
\email{egor.sobolev@gmail.com}
\affiliation{Institute of Mathematical Problems of Biology RAS, Pushchino, Russia}
\author{V.D.~Lakhno}
\email{lak@impb.psn.ru}
\affiliation{Institute of Mathematical Problems of Biology RAS, Pushchino, Russia}



\date{\today}

\begin{abstract}
The dynamics of charge migration
was modeled to calculate temperature dependencies of its thermodynamic equilibrium values
such as energy
and electronic heat capacity in homogeneous adenine fragments. The energy varies from nearly polaron one at $T \sim 0$
to midpoint of the conductivity band at high temperatures.
The peak on the graph of electronic heat capacity is observed at the polaron decay temperature.
\end{abstract}



\maketitle

\section{Introduction}

The problem of how a charge, or energy is transferred in molecular chains is of great interest for biology and nanobioelectronics --
a new rapidly developing discipline which combines achievements of nanoelectronics and molecular biology.
One of the central problems
of nanobioelectronics deals with creation of molecular nanowires and their circuits based on synthesized nucleotide chains.
Therefore the study of conducting properties of DNA molecules is of profound interest \cite{bib1,bib2,bib3}.

In this paper we carry out a direct numerical experiment to investigate temperature dependencies of
thermodynamic equilibrium values of ``electronic part of total energy'' and
electronic heat capacity in quasi-one-dimensional molecular chains.
The computations have
simulated homogeneous adenine-thymine polynucleotide sequences with excess charge carriers moving along adenine chain (polyA).

The knowledge of thermodynamic quantities is essential for understanding many fundamental properties of materials.
An important characteristic of all kinds of materials is their heat capacity.
For DNA nanobioelectronics, of primary concern is calculation of electronic part of the polynucleotide chains heat capacity.
Measurements of temperature dependence of electronic heat capacity gives the knowledge
of the concentration of current carriers in a molecule, quantum phase transitions, structural properties of materials, etc.
In the context of innovative development of high-accuracy calorimetry and nanocalorimetry \cite{bib4,bib5,bib6}
calculation of temperature dependencies of electronic heat capacities of homogeneous chains is a topical problem.

The model is based on Holstein Hamiltonian for charge transfer along a chain
\cite{bib7,bib8,bib9,bib10,bib11,bib12}:
\begin{align}
\hat{H} &=
\sum_{m,n} \nu_{nm} |m \rangle \langle n| +
\sum_n \chi_n u_{n} |n \rangle \langle n| +
\nonumber \\ &{}+
\frac 12 \sum_n M \dot{u}_{n}^2 +
\frac 12 \sum_n K u_{n}^2 .
\label{eq1}
\end{align}
Here $\nu_{mn}$ ($m\neq n$) are matrix elements of the electron transition
between $m$-th and $n$-th sites (depending on overlapping integrals),
$\nu_{nn}$ is the electron energy on the $n$-th site,
$\chi_n$ is a constant of electron coupling with displacements $u_n$ of the $n$-th site from the equilibrium position,
$M$ is the site's effective mass, $K$ is an elastic constant.
Choosing the wave function $\Psi$ in the form
$\Psi = \sum_{n=1}^N b_n |n\rangle$,
where $b_n$ is the amplitude of the probability of charge
(electron or hole) occurrence on the $n$-th site ($n=1,{\ldots} ,N$,
$N$ is the chain length),
from Hamiltonian \eqref{eq1} for homogenous chain we get motion equations:
\begin{align}
i \hbar \dot{b}_n &=
\nu_{n,n-1} b_{n-1} + \nu_{n,n+1} b_{n+1} + \chi u_n b_n,
\label{eq2} \\
\ddot{u}_n &=
-\omega^2 u_n - \chi |b_n|^2 -
\gamma \dot{u}_n + A_n (t).
\label{eq3}
\end{align}
Classical subsystem \eqref{eq3} includes a term with friction ($\gamma$ is a friction coefficient)
and a random force $A_n(t)$ such that $\langle A_n (t) \rangle =0$,
$\langle A_n (t) A_m (t+s) \rangle = 2 k_B T \gamma \delta_{nm} \delta (s)$
which simulate a thermostat with a temperature $T$. Such a way of simulating the thermostat temperature
by Langevin equations \eqref{eq3} is well-known \cite{bib13,bib14,bib15}.

The model parameters corresponding to nucleotide pairs are \cite{bib11,bib16,bib17,bib180,bib181,bib182,bib183,bib184,bib185,bib187,bib8}:
$M = 10^{-21}$\,g, terahertz frequency of sites oscillations $\omega =10^{12}\, \text{sec}^{-1}$ corresponds to
rigidity of hydrogen bonds $K \approx 0.062$\,eV/\AA$^2$, the coupling constant $\chi = 0.13$\,eV/\AA.
In homogenous adenine chain the matrix element of the transition between neighbor sites is $\nu_{n,n\pm1} = -\nu$, $\nu = 0.030$\,eV.

For a given temperature, we computed set of samples (trajectories of system \eqref{eq2},\eqref{eq3}
from various initial data and with various generated ``random'' time-series) and calculated the time dependence
of the total energy $\langle E_{tot}(t) \rangle$  averaged over samples
(where in individual sample $  E_{tot}(t) = \langle \Psi | \hat{H} |\Psi \rangle$.
Individual trajectories were calculated in two ways: by 2o2s1g-method \cite{bib19}
and with matrix exponential \cite{bib20} where forced renormalization was added
(the total probability of charge's occurrence in the system must be equal to 1, so
variables $b_n$ are ``corrected'' so that $\sum |b_n (t)|^2 = 1$).

Usually we dealt with two sets, each including 100 samples. The first set is calculated from
polaron initial data corresponding to the lowest energy $E_{pol}$.
The second one -- from the uniform distribution $|b_n(t=0)|^2 = 1/N$, for which
the velocity and displacement of sites were determined from thermodynamic equilibrium distribution for a given temperature.
The second variant of the initial data at $T=0$ corresponds to the highest energy of the system.
Calculations were carried out for large time intervals (for more detail, see \cite{bib21}),
until the graphs of time dependencies $\langle E_{tot}(t) \rangle$ obtained for each set of samples
became similar. Then the equilibrium value of  $\langle E_{tot}(T) \rangle$ was calculated for a given thermostat temperature $T$.

Having found $\langle E_{tot}(T) \rangle$, we made a numerical estimate of the electronic part of the total energy
$E_e = \langle E_{tot}(T) \rangle - Nk_BT$, i.e.\ we subtracted the heat energy of the N chain oscillators
from the total energy of the system.
To assess the heat capacity $C_V$ we used two methods so that to check the quality of the calculated ensemble
from which the averages were found:
\begin{align*}
C_V = \frac{\partial \langle E_{tot}(T) \rangle}{\partial T} =
\frac{1}{k_B T^2} \big( \langle E_{tot}^2 (T) \rangle - \langle E_{tot}(T) \rangle^2 \big).
\end{align*}
The results obtained by these two procedures are close.
The electronic heat capacity $C_e$ was calculated by formulae $C_e = \partial E_e/ \partial T = C_V - Nk_B$.

\begin{figure}[htb]
\includegraphics[width=0.45\textwidth]{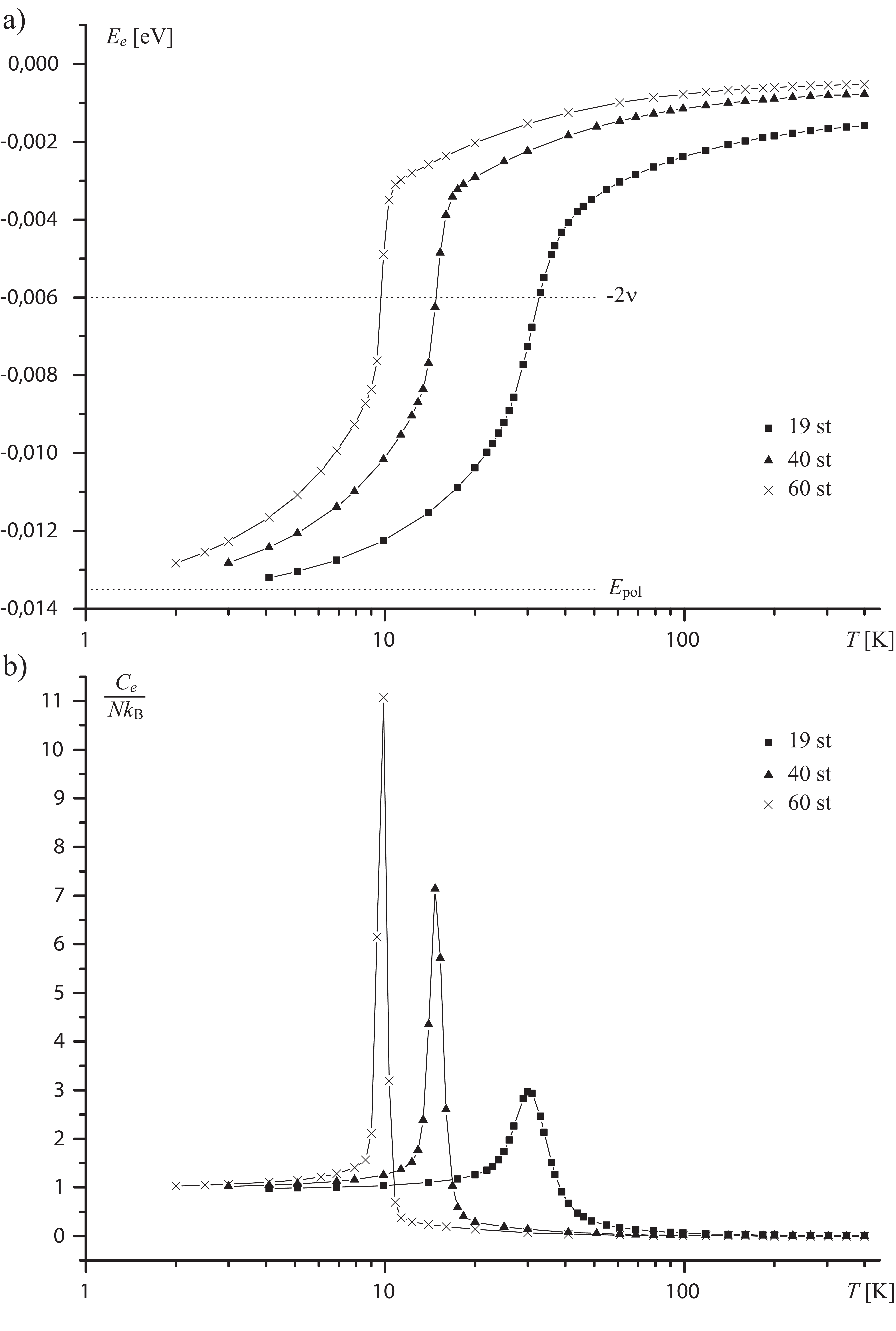}
\caption{Results of calculations of thermodynamically equilibrium values.
a) Electronic part of the total energy $E_e(T)$ for chains of length 19, 40 and 60 sites.
Dashed lines show polaron energies $E_{pol}$ and the lower bound of the conductivity band $2\nu$.
b) Normalized electronic heat capacity.
\label{fig1}}
\end{figure}

For various thermostat temperatures, we calculated the value of the total energy
in thermodynamic equilibrium $\langle E_{tot}(T) \rangle$ for polyA fragments of different lengths.
Fig.\ \ref{fig1} shows the results obtained for chains of 19, 40 and 60 sites.

Obviously, $\langle E_{tot}(T) \rangle$ depends on the chain length $N$.
The electronic part of the total energy $E_e(T)$ is also different for chains of different lengths.
However, the qualitative behavior is similar.
Fig.\ \ref{fig1}\,a shows the results of calculation of $E_e(T)$.
Fig.\ \ref{fig1}\,b demonstrates
the dependence of the ratio of $C_e$ and heat capacity of the  $N$ site chain on chain thermal energy $C_e / (N k_B)$.

For $T = 0$ $E_e = E_{pol}$, as the temperature grows, $E_e(T)$ increases and a charge passes on from polaron state to a delocalized one.
The polaron decay temperature depends not only on the model parameters but also on the chain length \cite{bib21}:
the larger is $N$, the less is the decay temperature. As the temperature grows still further, $E_e$ becomes a constant, which,
judging from calculations, is a function of the chain length $N$.
This constant is inside the conductivity band determined by eigenvalues of $W_k$ for a rigid chain (for $\chi =0$) \cite{bib22}:
$W_k = -2 \nu \cos \{k \pi /(N + 1)\}$, $k = 1,\dots,N$, and is close to zero,
i.e.\ the midpoint of the conductivity band, Fig.\ \ref{fig1}\,a.

Fig.\ \ref{fig1}\,b suggests that as the chain length increases, the peak $C_e / (N k_B)$ becomes higher and more sharp,
and shifts in respect of temperature toward zero.

As $T\to 0$, the normalized electronic heat capacity is independent of the chain length
and is close to heat capacity of a classical site $k_B$ (Fig.\ \ref{fig1}\,b).
Notice that the semiclassical approximation has meaning when the temperature is higher than the Debye one $\Theta$
($k_B \Theta = \hbar \omega$) \cite{bib15}.
For the DNA parameters used in this work $\Theta \approx 8$\,K.
Hence the peaks of electronic heat capacity for $N \ge 40$ (Fig.\ \ref{fig1}\,b) falls in the range of inapplicability
of the model under consideration. In short chains ($N < 20$) the electronic heat capacity peaks will be observed at temperatures $T > \Theta$,
however they are less pronounced. For $N = 19$ $\max C_e (T) \approx 57 k_B \approx 0.0049$\,eV/K for $T \approx 30$\,K,
for shorter chains the peak will be still lower.

\begin{figure}[htb]
\includegraphics[width=0.45\textwidth]{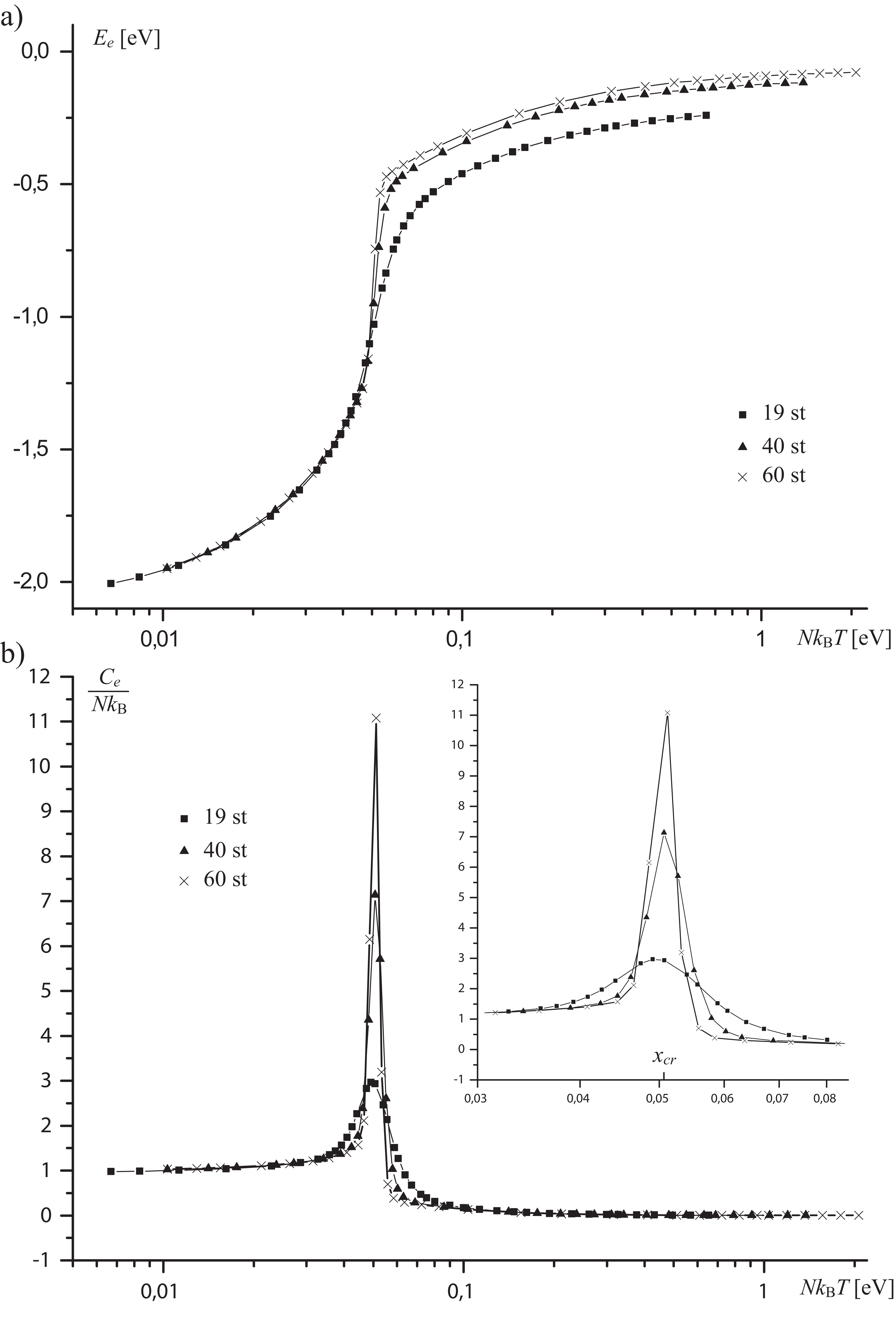}
\caption{Rescaled results of calculations.
a) Dependence of the charge energy $E_e$ on the energy of a classical chain.
b) Dependence of normalized electronic heat capacity $C_e/Nk_B$ on the energy of a classical
chain. The inset shows the peak of the function in more detail.
\label{fig2}}
\end{figure}

If we perform rescaling so that the variable $x$ be not the temperature, but the energy of a classical chain $E_{class} = k_B N T = x$,
then as is seen from Fig.\ \ref{fig2}\,à, the values of $E_e(x)$ for different $N$ are located in close proximity to one curve.
Fig.\ \ref{fig2}\,b shows the results of calculation of the normalized
electronic heat capacity $y(x) = C_e(x) / N k_B$.
On this scale the heat capacity peaks coincide at abscissa ($x \approx 0.05$) for different $N$. The decrease of $y(x)$
following the peak is nearly similar in all the cases. We approximated the values obtained by power law $y = {\mathit{Const}} \cdot x^b$
and found $b \approx -2$, i.e.\ when a charge in a chain is delocalized
the electronic heat capacity of a charge in a chain of $N$ sites decreases approximately as $(NT)^{-2}$.

The peaks of normalized heat capacity 
at $x_{cr} \approx 0.05$ (Fig.\ \ref{fig2}\,b)
correspond approximately to the energy value $E_e = -2\nu = -0.06$\,eV (bottom of the conductivity band) in Fig.\ \ref{fig2}\,a.

In our previous work \cite{bib21}, having calculated the delocalization parameter we showed that a polaron
in quasi-one-dimensional molecular chains has a characteristic decay temperature depending on the chain length.
Obviously, this polaron behavior should manifest itself in thermodynamic properties of the electronic part of the DNA heat capacity.
The peaks of the temperature dependence of the electronic heat capacity shown in Fig.\ \ref{fig2}
just fall into the range of polaron decay temperatures.
The peak at $x_{cr}$
corresponds to the transition between
two different types of state.
First is the polaron state with localized charge (Fig.\ \ref{fig2}\,b, left from the position of peak $x_{cr}$).
The second type is delocalized state, when the probability density is randomly distributed along the chain (Fig.\ \ref{fig2}\,b, right from $x_{cr}$).

The results of Fig.\ \ref{fig2}\,b seem paradoxical at low temperatures,
because at $T\to 0$ the value of the electronic heat capacity $C_e$
is equal to the ``macroscopic value'' $N k_B$ if $N \to \infty$.
Actually, the upper bound of the polaron state $x_{cr}$ is a constant energy, i.e.\ independent of $N$ and $T$.
Thus, the heating energy $N k_B T$ for transition from polaron state to delocalized one
is finite even for infinite chain and is equal to $x_{cr}$.

Based on the results obtained let us assess the possibility of measuring the electron part of the DNA heat capacity.
We will proceed from the fact that the present-day high-accuracy calorimetry enable one to measure the energy difference of about of attojoules
($ E = 10^{-18}$\,J) \cite{bib23}. If we assume that the characteristic energy of the polaron state in DNA
is $\sim 5\cdot 10^{-2}$\,eV ($\approx 8\cdot 10^{-21}$\,J) and one polaron occurs in a chain of length less than 100 sites,
then we get that for the above-mentioned accuracy of measurements, the number of polynucleotide fragments measured
by a calorimeter should be greater than 100. This condition is easily fulfilled for not too strongly diluted DNA solutions.
In other words, the temperature dependence of the electronic heat capacity can reasonably be measured experimentally.
Bearing in mind fast progress in improving the accuracy of calorimeters it is reasonable to expect that in the near future a new branch,
such as DNA nanocalorimetry will emerge.

An important field of application of electronic DNA nanocalorimetry is to use it for measuring the concentration
of current carriers in DNA chains playing the role of nanowires. Another essential trend is to use nanocalorimetry
for determining nucleotide sequences in mapping on the human genome \cite{bib24}.

In this paper we dealt with homogeneous chains. The existence of a temperature interval of polaron decay
is a general property which will take place in inhomogeneous chains too. One would expect that in inhomogeneous DNA sequences
there is also a peak of heat capacity. The location and shape of the peak in this case will be highly specific
to the location of an electron in a chain which makes possible noninvasive electronic way of mapping on the human genome.

\begin{acknowledgments}
The authors are thankful to the Joint Supercomputer Center of RAS for computer resources provided.
The work was done with partial support from the Russian Foundation for Basic Research,
projects No.\ 13-07-00256, 14-07-00894, 15-07-06426.
\end{acknowledgments}

\bibliography{refs}

\end{document}